\documentclass[12pt]{article}
\usepackage{amsmath}
\usepackage[psamsfonts]{amssymb}
\usepackage{cmmib57}
\usepackage[cmtip,arrow]{xy}
\usepackage{pb-diagram,pb-xy}
\newtheorem{Def}{\indent Definition}[section]
\newtheorem{Lem}[Def]{\indent Lemma}
\newtheorem{Prop}[Def]{\indent Proposition}
\newtheorem{Theo}[Def]{\indent Theorem}
\newtheorem{Cor}[Def]{\indent Corollary}
\newtheorem{Rem}[Def]{\indent Remark}
\newtheorem{Exa}[Def]{\indent Example}
\newtheorem{Pro}[Def]{\indent Problem}
\newcommand{\myskip}{\vspace*{8pt}}
\newcommand{\bDf}{\begin{Def}\em}
\newcommand{\eDf}{\end{Def}}
\newcommand{\bLm}{\begin{Lem}}
\newcommand{\eLm}{\end{Lem}}
\newcommand{\bPr}{\begin{Prop}}
\newcommand{\ePr}{\end{Prop}}
\newcommand{\bTh}{\begin{Theo}}
\newcommand{\eTh}{\end{Theo}}
\newcommand{\bCr}{\begin{Cor}}
\newcommand{\eCr}{\end{Cor}}
\newcommand{\bRm}{\begin{Rem}\em}
\newcommand{\eRm}{\end{Rem}}
\newcommand{\bEx}{\begin{Exa}\em}
\newcommand{\eEx}{\end{Exa}}
\newcommand{\bPb}{\begin{Pro}\em}
\newcommand{\ePb}{\end{Pro}}
\DeclareFontFamily{U}{UWCyr}{}
\DeclareFontShape{U}{UWCyr}{m}{n}{%
    <5> <6> <7> <8> <9>
    <10> <10.95> <12> <14.4> <17.28> <20.74> <24.88> wncyr10
    }{}
\DeclareMathAlphabet{\cyrm}{U}{UWCyr}{m}{n}

\newcommand{\mysec}[1]{\section{#1}}

\newcommand{\myssec}[1]{\subsection{#1}}

\newcommand{\bEq}{\begin{eqnarray}}
\newcommand{\eEq}{\end{eqnarray}}
\newcommand{\beq}{\begin{eqnarray*}}
\newcommand{\eeq}{\end{eqnarray*}}
\newcommand{\bCd}{\bEq\begin{CD}}
\newcommand{\eCd}{\end{CD}\eEq}
\newcommand{\bcd}{\beq\begin{CD}}
\newcommand{\ecd}{\end{CD}\eeq}
\newcommand{\ben}{\begin{enumerate}}
\newcommand{\een}{\end{enumerate}}
\newcommand{\btb}{\begin{tabbing}}
\newcommand{\etb}{\end{tabbing}}
\newcommand{\bPf}{\par\vspace*{-4pt}\indent{\sc Proof.}\enskip}
\newcommand{\ePf}{\medskip}
\def\QED{\hskip0.1em\hfill\null\ \null\nobreak\hfill\kern3pt\vbox{\hrule\hbox
     {\vrule\kern1pt\vbox{\kern1.7pt\hbox{$\scriptscriptstyle{QED}$}
      \kern0.2pt}\kern1pt\vrule}\hrule}}

\def\END{\hskip0.1em\hfill\null\ \null\nobreak\hfill\kern3pt\vbox{\hrule\hbox
     {\vrule\kern1pt\vbox{\kern1.7pt\hbox{$\,\,\,\vspace{5pt}$}
      \kern0.2pt}\kern1pt\vrule}\hrule}}
\newcommand{\ie}{{\em i.e$.$ }}
\newcommand{\eg}{{\em e.g$.$ }}
\newcommand{\R}{I\!\!R}

\newcommand{\mto}{\mapsto}

\newcommand{\der}{\partial}

\DeclareMathOperator{\im}{im}

\DeclareMathOperator{\byd}{{\raisebox{.1ex}{:}{=}}}

\newcommand{\sub}{\subset}

\newcommand{\wed}{\wedge}

\newcommand{\com}{\!\circ\!}
\newcommand{\con}{\,\lrcorner\,}
\newcommand{\ten}{\!\otimes\!}

\newcommand{\ucar}[1]{\underset{#1}{\times}}

\newcommand{\owed}[1]{\overset{#1}{\wedge}}

\newcommand{\uset}[2]{\underset{#1}{#2}}

\newcommand{\olin}[1]{\overline{#1}}

\newcommand{\alp}{\alpha}
\newcommand{\bet}{\beta}
\newcommand{\gam}{\gamma}
\newcommand{\del}{\delta}
\newcommand{\eps}{\epsilon}

\newcommand{\lam}{\lambda}
\newcommand{\sig}{\sigma}

\newcommand{\Gam}{\Gamma}

\newcommand{\vartht}{\vartheta}

\newcommand{\bG}{\boldsymbol{G}}

\newcommand{\bK}{\boldsymbol{K}}

\newcommand{\bP}{\boldsymbol{P}}

\newcommand{\bT}{\boldsymbol{T}}
\newcommand{\bU}{\boldsymbol{U}}

\newcommand{\bX}{\boldsymbol{X}}
\newcommand{\bY}{\boldsymbol{Y}}

\newcommand{\cC}{\mathcal{C}}

\newcommand{\cE}{\mathcal{E}}
\newcommand{\cF}{\mathcal{F}}

\newcommand{\cL}{\mathcal{L}}

\newcommand{\cS}{\mathcal{S}}

\newcommand{\balp}{\boldsymbol{\alp}}

\renewcommand{\rfloor}{\con}
\def\con{{\offinterlineskip\lower 1truept\hbox{\kern2pt
\vbox to7truept{\vfill\hbox to4truept{\hrulefill}}\vrule \kern3pt}}}
\newcommand{\D}{\cyrm{D}}

\newcommand{\For}[1]{\overset{#1}{\Lambda}}
\newcommand{\Con}[1]{\overset{#1}{\cal{C}}}
\newcommand{\Hor}[1]{\overset{#1}{\cal{H}}}
\newcommand{\Var}[1]{\overset{#1}{\cal{V}}}
\newcommand{\Thd}[1]{\overset{#1}{\Theta}}

\newcommand{\Z}{\mathbb{Z}}
\newcommand{\bdg}{\begin{diagram}}
\newcommand{\edg}{\end{diagram}}
\title{{\Large {\bf Conservation laws for non--global Lagrangians}}}
\bigskip
\author{{\large A. Borowiec\thanks{Supported by KBN grant 2 P03B
144 19.},
M. Ferraris,
M. Francaviglia\thanks{Both of them
supported by GNFM of INdAM, MURST and University of
Torino.}\, and M. Palese\thanks{Supported by GNFM of INdAM, MURST
University of Torino and CNR
grant n. 203.01.71/03.01.02.}}}
\date{}
\begin{document}

\maketitle

\begin{abstract}

  In the Lagrangian framework for symmetries and conservation laws of
field theories, we investigate globality properties of conserved
currents associated with non--global Lagrangians admitting global
Euler--Lagrange morphisms. Our approach is based on the recent
geometric formulation of the calculus of variations on finite
order jets of fibered manifolds in terms of variational sequences.

\myskip


\end{abstract}

\mysec{Introduction}\label{1}

In the Lagrangian framework for symmetries and conservation laws
of field theories, we investigate globality properties of
conserved currents associated with non--global Lagrangians which
admit global Euler--Lagrange morphisms 
(see also \cite{Sar03}). Our approach is based on the geometric formulation
of the calculus of variations on finite order jets of fibered manifolds in terms of
variational sequences
\cite{Kru99}. It was shown in \cite{FPV98a} that the Lie
derivative operator with respect to fiber--preserving vector
fields passes to the quotient, thus yielding a new operator on the
sheaves of the variational sequence, which was called the {\em
variational Lie derivative\/}. Making use of a representation
given in \cite{Vit98} for the quotient sheaves of the variational
sequence as concrete sheaves of forms, some abstract versions of
Noether's theorems have been provided, which can be interpreted in
terms of conserved currents for Lagrangians and Euler--Lagrange
morphisms.

Non--global Lagrangians are here defined as \v Cech cochains
valued into the sheaf of generalized Lagrangians. We 
relate
globality properties to the topology of the relevant manifold in
terms of the \v Cech cohomology of the manifold with values in the
sheaves of the variational sequence (see also \cite{Bor92}). To
this aim we provide a slightly modified version of some well known
results due to \cite{AnDu80} (Theorems \ref{I} and \ref{II}). We
shall in particular investigate the case of \v Cech cochains of
Lagrangians admitting global Euler--Lagrange morphisms but having
non--trivial cohomology class. In this case globality properties
still hold true for the conserved quantities associated with the
cochain of Lagrangians itself. For analogous results obtained in a
different framework we refer the reader to the interesting paper by Aldrovandi \cite{Al02}.

In Section \ref{2} we state the main notation and recall some
basic facts about sheaves of forms on finite order jets of fibered
manifolds,
   together with some standard results about \v Cech cohomology.
In Subsection \ref{3} we recall general results concerning
symmetries in variational sequences. Section \ref{4} is concerned
with the main results of the paper. We prove the existence of
global conserved quantities associated with Lagrangian symmetries
and generalized Lagrangian symmetries of \v Cech cochains of
Lagrangians.

\mysec{Preliminaries and notation}\label{2}

\myssec{Sheaves of forms on jets of fibered manifolds}

Let us consider a fibered manifold $\pi : \bY \to \bX$, with $\dim
\bX = n$ and $\dim \bY = n+m$. For $r \geq 0$ we are concerned
with the $r$--jet space $J_r\bY$ of jet prolongations of sections of 
the fibered manifold $\pi$; in particular, we set $J_0\bY
\equiv \bY$. We recall the natural fiberings $\pi^r_s : J_r\bY \to
J_s\bY$, $r \geq s$, and $\pi^r : J_r\bY \to \bX$; among these the
fiberings $\pi^r_{r-1}$ are {\em affine bundles}.

Greek indices $\lam ,\mu ,\dots$ run from $1$ to $n$ and they
label basis coordinates, while Latin indices $i,j,\dots$ run from
$1$ to $m$ and label fibre coordinates, unless otherwise
specified. We denote multi--indices of dimension $n$ by boldface
Greek letters such as $\balp = (\alp_1, \dots, \alp_n)$, with $0
\leq \alp_\mu$, $\mu=1,\ldots,n$; by an abuse of notation, we
denote with $\lam$ the multi--index such that $\alp_{\mu}=0$, if
$\mu\neq \lam$, $\alp_{\mu}= 1$, if $\mu=\lam$. We also set
$|\balp| \byd \alp_{1} + \dots + \alp_{n}$. The charts induced on
$J_r\bY$ are denoted by $(x^\lam,y^i_{\balp})$, with $0 \leq
|\balp| \leq r$; in particular, we set $y^i_{\bf{0}} \equiv y^i$. They 
are fibered charts, so that the choice of different letters ($x$ for 
the basis and $y$ for the fibers) stresses different transformation 
laws: in fact, fibered transformation laws of the kind $y'=y'(x,y)$ 
and $'=x'(x)$.
The local bases of vector fields and $1$--forms on $J_r\bY$
induced by the above coordinates are denoted by $(\der_\lam
,\der^{\balp}_i)$ and $(d^\lam,d^i_{\balp})$, respectively.

The {\em contact maps\/} on jet spaces induce the natural
complementary fibered morphisms over the affine fiber bundle
$J_r\bY \to J_{r-1}\bY$ \bEq\label{affine1} \D_{r} : J_r\bY
\ucar{\bX} T\bX \to TJ_{r-1}\bY \,, \qquad \vartht_{r} : J_r\bY
\ucar{J_{r-1}\bY} TJ_{r-1}\bY \to VJ_{r-1}\bY\,, \qquad r\geq 1\,,
\eEq with coordinate expressions, for $0 \leq |\balp| \leq r-1$,
given by \bEq\label{affine2} \D_{r} &= d^\lam\ten {\D}_\lam =
d^\lam\ten (\der_\lam + y^j_{\balp+\lam}\der_j^{\balp}) \,, \qquad
\vartht_{r} &= \vartht^j_{\balp}\ten\der_j^{\balp} =
(d^j_{\balp}-y^j_{{\balp}+\lam}d^\lam) \ten\der_j^{\balp} \,, \eEq
and the natural fibered splitting \cite{Kru99} \bEq \label{jet
connection} J_r\bY\ucar{J_{r-1}\bY}T^*J_{r-1}\bY =(
J_r\bY\ucar{J_{r-1}\bY}T^*\bX) \oplus\im \vartht_{r}^{*}\,. \eEq

The above splitting induces also a decomposition of the exterior
differential on $\bY$ in the {\em horizontal\/} and {\em vertical
differential\/}, $(\pi^{r+1}_r)^*\com\, d = d_H + d_V$.

A {\em projectable vector field} on $\bY$ is defined to be a pair
$(\Xi,\xi)$, where the vector field $\Xi:\bY \to T\bY$ 
is a fibered morphism over
the vector field $\xi: \bX \to T\bX$. By $(j_{r}\Xi, \xi)$ we denote the jet
prolongation of $(\Xi,\xi)$, and by $j_{r}\Xi_{H}$ and
$j_{r}\Xi_{V}$, respectively, the horizontal and the vertical part
of $j_{r}\Xi$ with respect to the splitting \eqref{jet
connection}.

i. For $r \geq 0$, we consider the standard sheaves $\For{p}_r$ of
$p$--forms on $J_r\bY$.

ii. For $0 \leq s \leq r $, we consider the sheaves
$\Hor{p}_{(r,s)}$ and $\Hor{p}_r$ of {\em horizontal forms\/}, \ie
of local fibered morphisms over $\pi^{r}_{s}$ and $\pi^{r}$ of the
type $\alp : J_r\bY \to \owed{p}T^*J_s\bY$ and $\bet : J_r\bY \to
\owed{p}T^*\bX$, respectively.

iii. For $0 \leq s < r$, we consider the subsheaf $\Con{p}_{(r,s)}
\sub \Hor{p}_{(r,s)}$ of {\em contact forms\/}, \ie of sections
$\alp \in \Hor{p}_{(r,s)}$ with values into $\owed{p}\,\im
\vartht_{s-1}^{*}$. There is a distinguished subsheaf $\Con{p}{_r}
\sub \Con{p}_{(r+1,r)}$ of local fibered morphisms $\alp \in
\Con{p}_{(r+1,r)}$ such that $\alp = \owed{p}\vartht_{r+1}^* \com
\Tilde{\alp}$, where $\tilde{\alp}$ is a section of the fibration
$J_{r+1}\bY \ucar{J_r\bY}$ $\owed{p}V^*J_r\bY$ $\to J_{r+1}\bY$
which projects down onto $J_{r}\bY$.

According to \cite{Vit98}, the fibered splitting \eqref{jet
connection} naturally yields the sheaf splitting
$\Hor{p}_{(r+1,r)}$ $=$ $\bigoplus_{t=0}^p$ $\Con{p-t}_{(r+1,r)}$
$\wed\Hor{t}_{r+1}$, which restricts to the inclusion $\For{p}_r$
$\sub$ $\bigoplus_{t=0}^{p}$
$\Con{p-t}{_r}\wed\Hor{t}{_{r+1}^{h}}$, where
$\Hor{p}{_{r+1}^{h}}$ $\byd$ $h(\For{p}_r)$ for $0 < p\leq n$ and
$h$ is defined to be the restriction to $\For{p}_{r}$ of the
projection of the above splitting onto the non--trivial summand
with the highest value of $t$.

Let $\alp\in\Con{1}_r\wed\Hor{n}{_{r+1}^h}$. Then there is a
unique pair of sheaf morphisms \bEq\label{first variation}
E_{\alp} \in \Con{1}_{(2r,0)}\wed\Hor{n}{_{2r+1}^{h}} \,, \qquad
F_{\alp} \in \Con{1}_{(2r,r)} \wed \Hor{n}{_{2r+1}^h} \,, \eEq
such that $(\pi^{2r+1}_{r+1})^*\alp=E_{\alp}-F_{\alp}$, and
$F_\alp$ is locally of the form $F_{\alp} = d_{H}p_{\alp}$, with
$p_{\alp} \in \Con{1}_{(2r-1,r-1)}\wed\Hor{n-1}{_{2r}}$ (see {\em
e.g.} \cite{Vit98}).

Recall (see \cite{Vit98}) that if
$\bet\in\Con{1}_{s}\wed\Con{1}_{(s,0)}\wed\Hor{n}_{s}$, then,
there is a unique $ \Tilde{H}_{\bet} \in
\Con{1}_{(2s,s)}\otimes\Con{1}_{(2s,0)}\wed\Hor{n}_{2s} $ such
that, for all $\Xi:\bY\to V\bY$, $ E_{\Hat{\bet}} = C^1_1 (
j_{2s}\Xi \ten \Tilde{H_{\bet}} ) $, where $\Hat{\bet} \byd
{j_{s}\Xi}\rfloor \bet$, $\rfloor$ denotes the inner product and
$C^1_1$ stands for tensor contraction. Then there is a unique pair
of sheaf morphisms \bEq\label{second} H_{\bet} \in
\Con{1}_{(2s,s)}\wed\Con{1}_{(2s,0)}\wed\Hor{n}_{2s} \,, \quad
G_{\bet} \in \Con{2}_{(2s,s)}\wed\Hor{n}_{2s} \,, \eEq such that
${\pi^{2s}_{s}}^*\bet=H_{\bet} - G_{\bet}$ and $H_{\bet} =
\frac{1}{2} \, A(\Tilde{H}_{\bet})$, where $A$ stands for
antisymmetrisation. Moreover, $G_\bet$ is locally of the type
$G_{\bet} = d_H q_{\bet}$, where $q_{\bet} \in
\Con{2}_{2s-1}\wed\Hor{n-1}_{2s-1}$, hence $[\bet] = [H_{\bet}]$.
Coordinate expressions of the morphisms $E_{\alp}$ and $H_{\bet}$
can be found in \cite{Vit98}. The morphism $\Tilde{H}$ is called
the Helmhotz--Sonin morphism associated with an Euler--Lagrange
type morphism. It is a global morphism the kernel of which
expresses the Helmholtz conditions for a given Euler--Lagrange
type morphism to be locally variational, \ie $\eta=E_{d_{V}\lam}$
\cite{Vit98}.

\myssec{\v Cech cohomology}

Suppose $\mathfrak{H}$ is a (paracompact Hausdorff) topological space.
In the following we shall call {\em graded sheaf} over $\mathfrak{H}$
any countable family of sheaves
$\cF^{*}\byd\{\cF^{i}\}_{i\in \Z}$ over $\mathfrak{H}$.
A {\em resolution} of a given sheaf $\cS$
is an exact sequence of sheaves of the form
$0\to\cS\to\cF^{*}$.

Set $H^{q}(\mathfrak{H},\cS )\byd \ker (\cC^{q}(\cS )_{\mathfrak{H}}
\to\cC^{q+1}(\cS )_{\mathfrak{H}})$ $/$
$\im (\cC^{q-1}(\cS )_{\mathfrak{H}}\to\cC^{q}(\cS )_{\mathfrak{H}})$,
for each $q\in \Z$, with $\cC^{-1}(\cS )_{\mathfrak{H}}\byd 0$. Here
$\cC^{q}(\cS )_{\mathfrak{H}}$ is the sheaf naturally induced by the
sheaf of discontinuous sections of $\cS$ \cite{Bre67}.
The Abelian group $H^{q}(\mathfrak{H},\cS )$ is called the cohomology group
of $\mathfrak{H}$ of degree $q$ with coefficients in the sheaf $\cS$.

We say $H^{*}(\mathfrak{H},\cS )\byd\oplus_{i\in \Z}
H^{i}(\mathfrak{H},\cS )$ to be the {\em cohomology of $\mathfrak{H}$
with values in} $\cS$.
It is clear that $H^{0}(\mathfrak{H},\cS )=\cS_{\mathfrak{H}}$.
We say $\cS$ to be {\em acyclic} if $H^{q}(\mathfrak{H},\cS )=0$ for
all $q\in \Z$,
$q>0$.

We remark that a resolution $0 \to\cS \to\cF^{*}$ naturally
induces a cochain complex $0 \to
\cS_{\mathfrak{H}}\to\cF^{*}_{\mathfrak{H}}$ via the global
section functor. Hence, we can define the {\em derived groups}
$H^{q}(\cF^{*}_{\mathfrak{H}} )\byd\newline \ker
(\cF^{q}_{\mathfrak{H}}\to\cF^{q+1}_{\mathfrak{H}})/\im
(\cF^{q-1}_{\mathfrak{H}} \to\cF^{q}_{\mathfrak{H}})$, for all
$q\in \Z$, with $\cF^{-1}_{\mathfrak{H}}\byd\cS_{\mathfrak{H}}$.

Let $0 \to \cS\to\cF^{*}$ be a resolution of $\cS$.
Then for each $q\in \Z$ there is a natural
morphism $H^{q}(\cF^{*}_{\mathfrak{H}})\to H^{q}(\mathfrak{H},\cS )$.
If the sheaves of $\cF^{*}$ are acyclic then the above morphism is an
isomorphism (Abstract de Rham Theorem) \cite{Bre67}.

We also recall that a {\em cochain complex\/}
is a sequence of morphisms of Abelian groups of the form
$0 \to \For{0}\to^{d_{0}} \For{1}\to^{d_{1}} \For{2}\to^{d_{2}}\dots$,
such that $d_{k+1} \com d_{k} = 0$. This last condition is equivalent to
$\im d_{k} \sub \ker d_{k+1}$. A cochain complex is said to be an {\em
exact sequence\/} if $\im d_{k} = \ker d_{k+1}$.

\myskip

Suppose now that $\cS$ is a sheaf of
Abelian groups over $\mathfrak{H}$. Let $\mathfrak{U}\byd\{U_{i}\}_{i\in I}$,
with $I\sub \Z$, be a countable
open covering of $\mathfrak{H}$. We set $C^{q}(\mathfrak{U},\cS )$ to be
the set of $q$--cochains with
coefficients in $\cS$. Let $\sig
=(U_{i_{0}},\dots ,U_{i_{q+1}})\sub \mathfrak{U}$ be a $q$--simplex and
$f\in C^{q}(\mathfrak{U},\cS )$.
The {\em coboundary operator} $\mathfrak{d}: C^{q}(\mathfrak{U},\cS )\to
C^{q+1}(\mathfrak{U},\cS )$ is the map defined by
\beq
\mathfrak{d} f(\sig )\byd \sum^{q+1}_{i=0}(-1)^{i}r^{|\sig_{i}|}_{|\sig |}
f(\sig_{i})\,,
\eeq
where $\sig_{j}\byd (U_{i_{0}},\dots ,U_{i_{j-1}},U_{i_{j+1}},\dots
U_{i_{q+1}})$, for $0\leq j\leq q+1$, $r$ is the restriction mapping of
$\cS$ and $|\sig_{j}|$ denotes the lenght of $\sig_{j}$
(see \cite{BoTu82}).

For all $q\in \Z$ the set $C^{q}(\mathfrak{U},\cS )$ can be endowed
with an Abelian group structure in a natural way. It is rather easy to verify
that $\mathfrak{d}$
is a group morphism, such that $\mathfrak{d} ^{2}=0$.
Hence we have the cochain complex
$
C^{0}(\mathfrak{U},\cS )\to C^{1}(\mathfrak{U},\cS )\to
C^{2}(\mathfrak{U},\cS )
\to\dots
$
\bDf
We say the derived groups $H^{*}(\mathfrak{U},\cS )$ of the above cochain
complex to be the {\em \v Cech cohomology} of the covering $\mathfrak{U}$ with
coefficients in $\cS$.\END
\eDf

The above cohomology is a combinatorial object and it depends on the choice
of a covering $\mathfrak{U}$.
Let $\mathfrak{U}\byd\{U_{i}\}_{i\in I}$, $\mathfrak{V}\byd\{V_{j}\}_{j\in J}$,
with $I,J\sub \Z$, be two countable coverings of $\mathfrak{H}$. Then we say
that $\mathfrak{V}$ is a {\em refinement} of $\mathfrak{U}$ if there
exists a map
$f:J\to I$ such that $V_{j}\sub U_{f(j)}$.
Then there is a group morphism
$H^{*}(\mathfrak{U},\cS )\to H^{*}(\mathfrak{V},\cS )$, so that we
can define the {\em \v Cech cohomology} of $\mathfrak{H}$
with coefficients in $\cS$ to be the direct limit
$H^{*}(\mathfrak{H},\cS )\byd \uset{\mathfrak{U}}{\lim}
H^{*}(\mathfrak{U},\cS)$.

\myssec{Cohomology of the variational sequence}\label{3}

We recall now the theory of variational sequences on finite order
jet spaces, as it was developed by Krupka \cite{Kru99}. By an
abuse of notation, we denote by $d\ker h$ the sheaf generated by
the presheaf $d\ker h$. Set $\Thd{*}_{r}$ $\byd$ $\ker h$ $+$
$d\ker h$.

\bDf The quotient sequence
\beq
0\arrow{e} \R_{\bY} \arrow{e} \dots\,\
\arrow[4]{e,t}{\cE_{n-1}}\,\ \ \For{n}_r/\Thd{n}_r
\arrow[3]{e,t}{\cE_{n}}\,\ \For{n+1}_r/\Thd{n+1}_r
\arrow[4]{e,t}{\cE_{n+1}}\,\ \ \For{n+2}_r/\Thd{n+2}_r
\arrow[4]{e,t}{\cE_{n+2}}\,\ \ \dots\,\ \arrow{e,t}{d} 0
\eeq
is called the $r$--th order {\em variational sequence\/}
associated with the fibered manifold $\bY\to\bX$. It turns out
that it is an exact resolution of the constant sheaf $\R_{\bY}$
over $\bY$ \cite{Kru99}.\END \eDf

\myskip

Let us now consider the cochain complex
\bEq\label{cochain}
0 \arrow{e} \R_{\bY} \arrow{e}  \dots\,\
\arrow[4]{e,t}{\cE_{n-1}}\,\ \ (\For{n}_r/\Thd{n}_r)_{\bY}
\arrow[3]{e,t}{\cE_{n}}\,\ \  (\For{n+1}_r/\Thd{n+1}_r)_{\bY}
\arrow[4]{e,t}{\cE_{n+1}}\,\ \  (\For{n+2}_r/\Thd{n+2}_r)_{\bY}
\arrow[4]{e,t}{\cE_{n+2}}\,\ \ \dots\,\
\arrow{e,t}{d} 0 
\eEq and denote by $H^k_{\text{VS}}(\bY)$
its $k$--th cohomology group. The variational sequence is a soft
resolution of the constant sheaf $\R_{\bY}$ over $\bY$, hence the
cohomology of the sheaf $\R$ is naturally isomorphic to the
cohomology of the cochain complex above. Also, the de Rham
sequence gives rise to a cochain complex of global sections, the
cohomology of which is naturally isomorphic to the cohomology of
the sheaf $\R_{\bY}$ on $\bY$, as an application of the Abstract
de Rham Theorem. Then, by a composition of isomorphisms, for all
$k\geq 0$ we get a natural isomorphism
$H^k_{\text{VS}}(\bY)\simeq H^k_{\text{dR}}\bY$ \cite{Kru99}.

\myskip

The quotient sheaves in the variational sequence can be conveniently
represented \cite{Vit98}. The sheaf morphism $h$ yields the natural
isomorphisms
\beq\label{isomorphisms}
& & I_{k} : \For{k}_{r}/\Thd{k}_{r} \to \Hor{k}{_{r+1}^{h}}\byd \Var{k}_{r} :
[\alp] \mto h(\alp)\,, \qquad k\leq n \,,
\\
& & I_{k} : (\For{k}_{r}/\Thd{k}_{r})\to
(\Con{k-n}{_{r}}\wed\Hor{n}{_{r+1}^h}) \big / h(\olin{d\ker h})\byd
\Var{k}_{r}: [\alp ]\mto [h(\alp )]\,, \qquad k > n \,.
\eeq

\myskip

Let $s\leq r$. Then we have the injective sheaf
morphism (see \cite{Kru99}) $\chi^{r}_s :(\For{k}_s/\Thd{k}_s)\to
(\For{k}_{r}/\Thd{k}_{r}) :
[\alp ] \mto [{\pi^{r}_s}^*\alp ]$,
where $[\alp ]$ denotes the equivalence class of a form $\alp$ on $J_s\bY$.

\mysec{\v Cech cochains valued in the sheaves of the variational
sequence}\label{4}

We
are interested in the case in which the topology of
$\bY$ is non--trivial; in particular we shall be concerned with an
application of \v Cech cohomology to the cases
$H^{n+1}_{\text{dR}}\bY \neq 0$ and $H^{n}_{\text{dR}}\bY \neq 0$.

The following results hold true (see \eg \cite{AnDu80}, and
\cite{BBH91} Chap. II).

\bTh\label{I} 
Let us consider the variational sequence
\eqref{cochain} and let $\bK_{r}\byd \text{Ker}\,\, \cE_{n}$ and
$H^{1}(\bY, \bK_{r})$ be the first \v Cech cohomology group of
$\bY$ with values in $\bK_{r}$. Then the long exact sequence
obtained from the short exact sequence 
\beq 0
\arrow{e}\bK_{r}\arrow{e} \Var{n}_{r}\arrow[2]{e,t}{\cE_{n}}\, \
\cE_{n}(\Var{n}_{r})\arrow{e} 0 \eeq gives rise to the exact
sequence \beq 0 \arrow{e} \Gam(\bY,\bK_{r})\arrow{e}
\Gam(\bY,\Var{n}_{r})\arrow{e}
\Gam(\bY,\cE_{n}(\Var{n}_{r}))\arrow{e,t}{\del} H^{1}(\bY,
\bK_{r})\arrow{e} 0 \,. 
\eeq 
\eTh

\bTh\label{II} Let us consider the variational sequence
\eqref{cochain} and let $\bT_{r}\byd \text{Ker}d_{H}$ and
$H^{1}(\bY, \bT_{r})$ be the first \v Cech cohomology group of
$\bY$ with values in $\bT_{r}$. Then the long exact sequence
obtained from the short exact sequence \beq 0 \to\bT_{r}\to
\Var{n-1}_{r}\arrow{e,t}{d_{H}} d_{H}(\Var{n-1}_{r})\to 0 \eeq
gives rise to the exact sequence \beq 0 \to\Gam(\bY,\bT_{r})\to
\Gam(\bY,\Var{n-1}_{r})\to
\Gam(\bY,d_{H}(\Var{n-1}_{r}))\arrow{e,t}{\del '} H^{1}(\bY,
\bT_{r})\to 0 \,. \eeq \eTh

Furthermore we have (\cite{AnDu80}, Lemma $4.1$, Theorem $4.2$,
\cite{Kru99}, \cite{Vit98}) for $s\leq r$:
\beq
H^{n+1}_{dR}(\bY)\simeq H^{n+1}_{VS}(\bY) \simeq H^{1}(\bY, \bK_{r})
\simeq H^{1}(\bY,\chi^{r *}_{s} \bK_{s})\,,
\eeq
and
\beq
H^{n}_{dR}(\bY)\simeq H^{n}_{VS}(\bY) \simeq H^{1}(\bY, \bT_{r})
\simeq H^{1}(\bY,\chi^{r *}_{s} \bT_{s})\,.
\eeq

\bRm\label{remark} As a straightforward application of the
Abstract de Rham Theorem, we have the following.

\myskip

Let $\eta\in(\Var{n+1}_{r})_{\bY}$ be a global section such that
$\cE_{n+1}(\eta)=0$. Suppose, moreover, that
$H^{n+1}_{\text{dR}}\bY \ni \del\eta = 0$. Then, there exists a
global section $\lam\in(\Var{n}_{r})_{\bY}$ such that
$\cE_n(\lam)=\eta$ (see \eg \cite{And86}).

\myskip

Analogously, let $\lam\in(\Var{n}_{r})_{\bY}$ be a global section
such that $\cE_{n}(\lam)=0$, \ie $\lam$ is variationally trivial.
Suppose, moreover, that $H^{n}_{\text{dR}}\bY \ni \del' \lam= 0$.
Then, there exists a global section $\bet\in(\Var{n-1}_{r})_{\bY}$
such that $\cE_{n-1}(\bet)=\lam$, where $\cE_{n-1}=d_{H}$ (see \eg
\cite{Kru99,Vit98}). \END \eRm If the topology of $\bY$ is
trivial, so that, in particular, $H^{n+1}_{\text{dR}}\bY =0$ and
$H^{n}_{\text{dR}}\bY = 0$ hold true, then each global
Euler--Lagrange morphism $\eta$ is globally variational and each
global variationally trivial Lagrangian $\lam$ is the horizontal
differential of a form $\bet$.

If the topology of $\bY$ is non--trivial, \ie
$H^{n+1}_{VS}(\bY) \simeq H^{1}(\bY, \bK_{r}) \neq 0$,
then the inverse problem for a given global Euler--Lagrange
morphism $\eta$ can be solved only locally, so that in general we can write
$\eta= \cE_{n}(\lam)$ only locally (provided, of course, that the corresponding
cohomology class of $\eta$ is non--trivial). More precisely this
means that around
each point a Lagrangian
$\lam_{\bU}$ is defined only on an open
subset $\bU\sub\bY$, so that $\eta|_{\bU}=\cE_{n}(\lam_{\bU}).$ We
are then naturally faced with the following situation which is in
fact often encountered in physical applications: there exists a
countable open covering $\{\bU_{i}\}_{i\in \Z}$ in $\bY$ together
with a family of local Lagrangians $\lam_{i}$ over each subset
$\bU_{i}\sub\bY$ (which, a priori, do not glue together into a
global Lagrangian $\lam$). Let then $\mathfrak{U}\byd \{\bU_{i}\}_{i\in I}$,
with $I\sub\Z$, be any countable open covering of $\bY$ and
$\lam= \{\lam_{i}\}_{i\in I}$ a $0$--cochain of Lagrangians in \v Cech
cohomology with values in the sheaf $\Var{n}_{r}$, \ie $\lam\in
C^{0}(\mathfrak{U},\Var{n}_{r})$. By an abuse of notation we shall
denote by $\eta_{\lam}$ the $0$--cochain formed by the restrictions
$\eta_{i}=\cE_{n}(\lam_{i})$.

\bRm Let $\mathfrak{d}\lam
=\{\lam_{ij}\}=(\lam_{i}-\lam_{j})|_{U_{i}\cap U_{j}}$. We stress
that $\mathfrak{d}\lam =0$ if and only if $\lam$ is globally
defined on $\bY$. Analogously, if $\eta \in
C^{0}(\mathfrak{U},\Var{n+1}_{r})$, then $\mathfrak{d}\eta =0$ if
and only if $\eta$ is global.\END \eRm

\bRm Let $\lam\in C^{0}(\mathfrak{U},\Var{n}_{r})$ and let
$\eta_{\lam}\byd \cE_{n}(\lam)\in
C^{0}(\mathfrak{U},\Var{n+1}_{r})$ be as above. Then
$\mathfrak{d}\lam=0$ implies $\mathfrak{d}\eta_{\lam}=0$, but the
converse is {\em not} true, in general. This is due to the
$\R$--linearity of all the operations involved in the variational
sequence. Therefore
$\mathfrak{d}\eta_{\lam}=\eta_{\mathfrak{d}\lam}=0$ implies only
$\mathfrak{d}\lam\in C^{1}(\mathfrak{U},\bK_r)$. \END \eRm

We shall in fact be concerned with the case
\bEq\label{nonglobal}
\mathfrak{d}\eta_{\lam}=0 \,, \qquad   \mathfrak{d}\lam \neq 0 \,.
\eEq

\bDf
We shall call a \v Cech cochain $\lam$ of Lagrangians satisfying condition
\eqref{nonglobal} a {\em non--global} Lagrangian.\END
\eDf

\bDf A non--global Lagrangian is said to be {\em topologically
non--trivial} if the cohomology class of $\eta_{\lam}$ in the
first \v Cech cohomology group is non--trivial, \ie
$\del\eta_{\lam}\neq 0$.\END \eDf

It is clear that a non--global Lagrangian
is defined modulo a refinement of $\mathfrak{U}$. In
particular, $\mathfrak{U}$ can be chosen to be a good
covering of $\bY$ (on a differentiable manifold there exists always a
good covering, see \eg
\cite{BoTu82}), on which the cohomology is trivial. Then Remark
\ref{remark} can be
reformulated as follows.

\bPr\label{trivial}
\noindent (A) Let $\lam\in\Var{n}_r$ be a global
variationally trivial Lagrangian. Then for any
good cover $\mathfrak{U}$ there exists a $0$--cochain $\bet\in
C^0(\mathfrak{U}, \Var{n-1}_r)$  such that $\lam=d_H\bet$.
Thus $\mathfrak{d}\bet\in C^1 (\mathfrak{U}, \bT_r)$ defines a
unique cohomology class
$[\mathfrak{d}\bet]_{\check{C}}\byd\del^\prime\lam\in H^1
(\bY, \bT_r)\simeq H^{n}_{dR}\bY$. If, moreover, this cohomology
class is trivial, then there exists a $0$--cochain $\gam\in C^0
(\mathfrak{U}, \Var{n-2}_r)$ such that
$d_H\mathfrak{d}\gam=\mathfrak{d}\bet$ with
$\bet^\prime=\bet-d_H\gam$ a global morphism and $\lam=d_H\bet^\prime$.

\noindent (B) Let $\lam\in C^{0}(\mathfrak{U},\Var{n}_{r})$ be a non--global
Lagrangian. Then $\mathfrak{d}\lam$ defines a unique cohomology
class $[\mathfrak{d}\lam]_{\check{C}}\byd\del\eta_\lam\in H^1 (\bY,
\bK_r)\simeq H^{n+1}_{dR}\bY$. If, moreover, this cohomology class
is trivial then there exists a $0$--cochain $\nu\in
C^0(\mathfrak{U},\Var{n-1}_{r})$ such that
\beq
\mathfrak{d}\lam=d_H\mathfrak{d}\nu\ \ \mbox{and}\ \
\mathfrak{d}(\lam-d_{H}\nu)=0\,.
\eeq
Thus $\lam^\prime=\lam-d_{H}\nu$ is a global Lagrangian and
$\cE_n(\lam^\prime)=\cE_n(\lam)$.
\ePr

\bPf It follows from the application of the
Poincar\'e Lemma, the standard \v Cech cohomology arguments
\cite{Bre67} and the Abstract de Rham Theorem (see also
\cite{AnDu80}; in \cite{KrMu98} was shown that, even more,
$\bet\in \cC^{0}(\mathfrak{U},\Var{n-1}_{r-1})$).\QED \ePf

\bEx (Einstein theory)
  From the above Proposition it follows that a topologically trivial
non--global Lagrangian is always equivalent to a global
one. This is \eg the case of the Hilbert--Einstein Lagrangian, which
is a second order global
Lagrangian in the bundle
$\bY=\text{Lor}(\bX)$ of Lorentzian metrics over $\bX$. The
Hilbert--Einstein Lagrangian is equivalent to
a sheaf of non--global first order Einstein's Lagrangians.
In this case, the cohomology class corresponding to
the Euler--Lagrange morphism is trivial and this fact does not depend on the
topology of the space-time manifold (see also \cite{FeFr91}).
\END
\eEx

\bEx (Chern-Simons theory) Let $\bP=\bP(\bX,\bG)$ be a principal
bundle over an odd dimensional manifold $\bX$ with a structure
group $\bG$ (\eg any simple Lie group). To any connection
one--form $\omega$ we can associate the Chern-Simons form
\cite{ChSi74} from which by pull-back along any (local) section
one gets a (local) Lagrangian on $\bX$. Since the Chern-Simons
form is not tensorial the local Lagrangians are not
gauge-invariant. In spite of this fact, the corresponding
Euler-Lagrange equations (\ie the vanishing curvature equations
for $\omega$) are invariant and global. Moreover, in this case, an
invariant Lagrangian does not exist at all. The existence of
global Lagrangians relies on the choice of a global section on
$\bP$ (see \eg \cite{BFF98,BFF01} and references quoted therein).
\END\eEx

\myssec{Symmetries and conservation laws}

Making use of the sheaf isomorphisms \eqref{isomorphisms} and of
the decomposition formulae
\eqref{first variation} and \eqref{second}, in
\cite{FPV98a} it was proved that the Lie derivative operator  with respect to
the $r$-th order prolongation $j_{r}\Xi$ of a  projectable vector field
$(\Xi,\xi)$ can be  conveniently represented on the quotient
sheaves of the variational  sequence in terms of an operator,
the {\em variational Lie derivative\/} $\cL_{j_{r}\Xi}$, as follows:

\noindent if $p = n$ and $\lam \in \Var{n}_{r}$, then
\bEq\label{Lieder} \cL_{j_{r}\Xi}\lam = \Xi_{V}\rfloor
\cE_{n}(\lam)+ d_{H}(j_{r}\Xi_{V}\rfloor p_{d_{V}\lam}+\xi\rfloor
\lam)\,;
\eEq
\noindent if $p = n+1$ and $\eta \in \Var{n+1}_{r}$,
then \bEq\label{helmholtz} \cL_{j_{r}\Xi} \eta=
\cE_{n}(\Xi_{V}\rfloor \eta)+\Tilde{H}_{d\eta}(j_{2r+1}\Xi_{V})
\,. \eEq

\bDf Let $(\Xi,\xi)$ be a projectable vector field on $\bY$. Let
$\lam \in \Var{n}_{r}$ be a Lagrangian and $\eta \in
\Var{n+1}_{r}$ an Euler--Lagrange morphism. Then $\Xi$ is called a
{\em symmetry\/} of $\lam$ (respectively, a {\em generalized or
Bessel--Hagen symmetry},  of
$\eta$) if
$\cL_{j_{r+1}\Xi}\,\lam = 0$ (respectively, if
$\cL_{j_{2r+1}\Xi}\,\eta= 0$). \END \eDf

Let now $\eta \in \Var{n+1}_{r}$ be an Euler--Lagrange morphism and let
$\sig : \bX \to \bY$ be a section. We recall that $\sig$ is said to be
{\em critical\/} if $\eta \circ j_{2r+1}\sig = 0$, \ie if it is a
solution of the
Euler--Lagrange equations $(j_{2r+1}\sig)^{*}\cE_{n}(\lam)=0$.

Let $\lam \in \Var{n}_{r}$ be a Lagrangian and $(\Xi,\xi)$
a symmetry of $\lam$. Then, by Equation \eqref{Lieder},
\ie the first Noether's theorem, we have
\beq
0 = \Xi_{V} \rfloor \cE_{n}(\lam) +
d_{H}(j_{r}\Xi_{V} \rfloor p_{d_{V}\lam}+ \xi \rfloor \lam)~.
\eeq
Suppose that the section $\sig:\bX \to \bY$ fulfils
$(j_{2r+1}\sig)^{*}(\Xi_{V} \rfloor \cE_{n}(\lam)) = 0$,
then we have the conservation law\,
$d ((j_{2r}\sig)^{*}
(j_{r}\Xi_{V} \rfloor p_{d_{V}\lam}+ \xi \rfloor \lam)) = 0$.

The above implies that
$\del^\prime(\cL_{j_{r}\Xi}\lam)\equiv\del^\prime(\Xi_{V}\rfloor
\eta_\lam)\equiv 0$.

\bDf Let $\lam \in \Var{n}_{r}$ be a Lagrangian and $\Xi$ a global
symmetry of $\lam$. Then a sheaf morphism of the type
$\eps(\lam,\Xi) = (j_{r}\Xi_{V} \rfloor p_{d_{V}\lam}+ \xi \rfloor
\lam)$  $\in$ $\Var{n-1}_{r}$ is said to be a {\em canonical} or
{\em Noether current\/}. \END \eDf

\bRm\label{arbitrary1}
Notice that if $\lam$ is globally defined,
then for any global symmetry $\Xi$ of $\lam$ the
morphism $\eps(\lam, \Xi)$ can be globally defined too (see \eg
\cite{Kol83}).
If $H^{n+1}_{\text{dR}}\bY \neq 0$,
\ie the topology of $\bY$ is not
trivial, then given a globally
defined Euler--Lagrange morphism with non--trivial cohomology class,
we cannot find a corresponding
globally defined Lagrangian via the inverse problem, so that in this case the
corresponding Noether conserved current $\eps$ is not global.
\END\eRm

\bRm
Let $\eta \in \Var{n+1}_{r}$ and let $\Xi$ be a generalized symmetry of $\eta$.
Then, by Equation \eqref{helmholtz}, we have
$0 = \cE_{n}(\Xi_{V} \rfloor \eta) +
\Tilde{H}_{d\eta}(j_{2r+1}\Xi_{V})$.
Suppose that $\eta$ is locally variational, \ie
$\Tilde{H}_{d\eta} = 0$; then we have\,
$\cE_{n}(\Xi_{V} \rfloor \eta) = 0$.
This implies that $\Xi_{V} \rfloor \eta$ is variationally trivial.
Therefore
$\Xi_{V} \rfloor \eta$ is locally of the type
$\Xi_{V} \rfloor \eta = d_{H}\bet$, where $\bet \in
\cC^{0}(\mathfrak{U},\Var{n-1}_{r+1})$ (in \cite{Gri99,KrMu98} was
shown that, even more,
$\bet \in
\cC^{0}(\mathfrak{U},\Var{n-1}_{r-1})$).

Suppose that the section $\sig:\bX \to \bY$ fulfils
$(j_{2r+1}\sig)^{*}(\Xi_{V} \rfloor \eta) = 0$. Then we have
$d ((j_{2r}\sig)^{*} \bet) = 0$
so that, as in the case of Lagrangians, if $\sig$ is critical,
then $\bet$ is conserved along $\sig$.
\END
\eRm

\bDf
Let $\eta \in \Var{n+1}_{r}$ be an  Euler--Lagrange morphism
and $\Xi$ a symmetry of $\eta$. Then a sheaf morphism of the type
$\bet$ fulfilling the conditions of the above
Remark is called a {\em generalized conserved current\/}.\END
\eDf

Notice that, for locally variational Euler--Lagrange morphisms, \ie
$\eta=\eta_\lam\equiv\cE_n(\lam)$ or, equivalently
$\Tilde{H}_{d\eta}(j_{2r+1}\Xi_{V})=0$. This implies
$\cL_{j_{2r}\Xi} \eta_\lam=
\cE_{n}(\Xi_{V}\rfloor \eta_\lam) = \cE_n(\cL_{j_{r}\Xi}\lam)$.

\bRm
Since $(j_{r}\sig)^{*}d_H\eps =d((j_{r}\sig)^{*}\eps)
= 0$ any solution $\sig$ defines a corresponding cohomology
class $\sig(\eps)\equiv [(j_{r}\sig)^{*}\eps]_{\check{C}}$ $\in$
$H^{n-1}_{dR}\bX$. If all these cohomology classes are trivial
then the corresponding current is called trivial (otherwise it is
called topological). It is obvious that currents admitting
(global) superpotentials \cite{FFP98} are trivial in the above sense.
Non--trivial currents are more interesting and lead to topological
charges (see e.g. \cite{Tor94}). Notice that if $H^{n-1}_{dR}\bX=0$ then
topological charges do not appear.\END\eRm

Due to $\cE_{n}\cL_{j_{r}\Xi} = \cL_{j_{2r+1}\Xi}\cE_{n}$,
a symmetry of a Lagrangian $\lam$ is also a symmetry of its
Euler--Lagrange morphism $\cE_n(\lam)$ but the converse is not
true. If  $(\Xi, \xi)$ is a generalized symmetry of $\lam$ the corresponding
current is not longer a canonical Noether conserved current for
$\lam$, in general.

Instead we can state the following (see \cite{Tra67,Tra96} for the
local version).

\bPr
Let $(\Xi,\xi)$ be a generalized symmetry for a (global) Lagrangian
$\lam$ $\in$ $\Var{n}_r$. Thus the canonical Noether current is
not conserved in general. If the cohomology class
$\del^\prime(\Xi_{V}\rfloor\eta_{\lam})$ $\in$ $H^n_{dR}\bY$ is
trivial then there exists a global conserved
  current associated with $(\Xi,\xi)$. \ePr

\bPf When $\cL_{j_{r}\Xi}\lam =0$ then we are in the standard
Noether case. If $\cL_{j_{r}\Xi}\lam\neq 0$ then
$\cL_{j_{r}\Xi}\lam$ is variationally trivial with the trivial
cohomology class $\del^\prime(\cL_{j_{r}\Xi}\lam)$ (Remark
\ref{arbitrary1}). Hence, there exits a global morphism
(Proposition \ref{trivial} (A)) $\beta\equiv \bet(\lam, \Xi)$ such
that $\cL_{j_{r}\Xi}\lam = d_H\bet(\lam, \Xi)$ and
\bEq\label{improved}  \Xi_{V}\rfloor \eta_\lam=d_{H}(\eps(\lam,
\Xi)-\bet(\lam, \Xi))\,. \eEq Thus $\tilde\eps(\lam,
\Xi)\byd\eps(\lam, \Xi)-\bet(\lam, \Xi)$ is global and conserved.
\QED \ePf

As a result it is then  possible to get a realization of the
corresponding conservation law associated with this generalized
symmetry, in terms of a (non--canonical) conserved current which
is global.

\bDf We call the above non--canonical conserved current an {\em
improved \newline Noether current}. \END\eDf

\bRm\label{arbitrary2} We stress that if $H^{n}_{\text{dR}}\bY =
0$, the improved Noether current $\tilde\eps(\lam, \Xi)$ is
always conserved and globally defined. If $H^{n}_{\text{dR}}\bY
\neq 0$ this is not true, in general. More precisely, for
$\del^\prime(\Xi_{V}\rfloor \eta_\lam)\neq 0$ we have $\bet(\lam,
\Xi)$ $\in$ $C^0(\mathfrak{U}, \Var{n-1}_r)$ with non--trivial
cohomolgy class $[\mathfrak{d}\bet(\lam,
\Xi)]_{\check{C}}=\del^\prime(\Xi_{V}\rfloor \eta_\lam)$ and therefore,
$\tilde\eps(\lam, \Xi)$ is conserved but not global. This is \eg the
case where topological charges can appear.\END \eRm

We are in a similar situation for a non--global Lagrangian
$\lam$ $\in$ $C^0(\mathfrak{U},\Var{n}_r)$. In this case both
components of $\tilde\eps(\lam, \Xi)$, the canonical and the improved one,
are non--global too. However, Equation \eqref{improved} still holds
true for $\tilde\eps(\lam, \Xi)$ $\in$
$C^0(\mathfrak{U},\Var{n-1}_r)$ with
$[\mathfrak{d}\tilde\eps(\lam, \Xi)]_{\check{C}}$ $=$
$\del^\prime(\Xi_{V}\rfloor\eta_{\lam})$ and it provides us
the following.

\bPr Let $(\Xi,\xi)$ be a global generalized symmetry for a
non--global Lagrangian $\lam$ $\in$ $C^0(\mathfrak{U},\Var{n}_r)$.
Then the improved Noether current is conserved and, in general,
non--global. It is possible to improve it further to a global
conserved current provided that
$\del^\prime(\Xi_{V}\rfloor\eta_{\lam})=0$. \ePr

\bPf Since $[\mathfrak{d}\tilde\eps(\lam, \Xi)]_{\check{C}}=0$ then, thanks to
Proposition \ref{trivial} (A), it can be globalized. \QED \ePf

\bRm
If $\del(\eta_\lam)\neq 0$, then
  $\del^\prime(\Xi_{V}\rfloor\eta_{\lam})=0$ is not true, in general.
Therefore, in
order to get a global conserved quantity for topologically
non--trivial Lagrangians some of our assumptions need to be relaxed.
For example, one could consider $0$--cochains of
symmetries instead of global projectable vector fields. This may
cover some physically interesting cases, like translations or
angular momentum, \eg. This will be the subject of our future
investigations.
\END\eRm

\subsection*{Acknowledgments}

\noindent Thanks are due to I. Kol\'a\v r, D. Krupka and R. Vitolo
for many valuable
discussions.


\noindent{\em Authors' addresses:}
\\{\footnotesize Institute of Theoretical Physics, University of Wroc{\l}aw}
\\{\footnotesize pl. Maxa Borna 9, 50-204  Wroc{\l}aw, Poland}.
\\{\footnotesize Department of Mathematics,
University of Torino}
\\{\footnotesize via C. Alberto 10, 10123 Torino, Italy}.
\\{\footnotesize e--mails: {\sc borow@ift.uni.wroc.pl,
ferraris@dm.unito.it,}} \\
{\footnotesize {\sc francaviglia@dm.unito.it, palese@dm.unito.it}}.

\end{document}